\documentclass[prl,twocolumn,superscriptaddress,
amsmath,amssymb,showpacs]{revtex4}

\usepackage{bm,graphicx,hyperref}

\newcommand{\rbd}{$^{87\!}{\rm Rb}$}
\newcommand{\csm}{$^{133\!}{\rm Cs}$}

\begin{document}

\title{Quantum Metrology: Dynamics vs.~Entanglement}

\author{Sergio Boixo}
\affiliation{Department of Physics and Astronomy, University of New
Mexico, Albuquerque, New Mexico 87131-0001, USA}
\affiliation{Los Alamos National Laboratory, Los Alamos, New Mexico 87545, USA}

\author{Animesh Datta}
\affiliation{Department of Physics and Astronomy, University of New
Mexico, Albuquerque, New Mexico 87131-0001, USA}

\author{Matthew J.~Davis}
\affiliation{School of Physical Sciences, University of Queensland,
Brisbane, Queensland 4072, Australia}

\author{Steven T.~Flammia}
\affiliation{Perimeter Institute for Theoretical Physics, 31 Caroline
Street North, Waterloo, Ontario N2L 2Y5, Canada}

\author{Anil Shaji}
\email{shaji@unm.edu}
\affiliation{Department of Physics and Astronomy, University of New
Mexico, Albuquerque, New Mexico 87131-0001, USA}

\author{Carlton M.~Caves}
\affiliation{Department of Physics and Astronomy, University of New
Mexico, Albuquerque, New Mexico 87131-0001, USA}
\affiliation{School of Physical Sciences, University of Queensland,
Brisbane, Queensland 4072, Australia}

\begin{abstract}
A parameter whose coupling to a quantum probe of $n$ constituents
includes all two-body interactions between the constituents can be
measured with an uncertainty that scales as $1/n^{3/2}$, even when
the constituents are initially unentangled. We devise a protocol that
achieves the $1/n^{3/2}$ scaling without generating any entanglement
among the constituents, and we suggest that the protocol might be
implemented in a two-component Bose-Einstein condensate.
\end{abstract}

% 03.65.Ta 	Foundations of quantum mechanics; measurement theory
% 03.65.-w 	Quantum mechanics
% 03.75.Nt 	Other Bose–Einstein condensation phenomena 
% 03.75.Mn 	Multicomponent condensates; spinor condensates 

\maketitle

Quantum mechanics determines the fundamental limits on measurement
precision.  In the prototypal quantum metrology scheme, the value of
a parameter is imprinted on a quantum probe through an interaction in
which the parameter appears as a coupling
constant~\cite{giovannetti06a}.  The number $n$ of constituents in
the probe is often considered to be the most important resource for
such schemes.  We denote the parameter to be estimated by $\gamma$,
and we write the interaction Hamiltonian as $\mathcal{H}=\hbar\gamma
H$, where $H$ is a dimensionless coupling Hamiltonian. The
measurement precision is quantified by the units-corrected
root-mean-square deviation of the estimate $\gamma_{\rm est}$ from
its true value,
\begin{equation}
\label{deltagamma}
\delta \gamma = \bigg \langle \bigg(  \frac{\gamma_{\rm est}}
{\partial\langle \gamma_{\rm est} \rangle / \partial \gamma }
- \gamma \bigg)^{\!2} \bigg\rangle^{\!1/2}.
\end{equation}

The essential point made
in~\cite{luis04a,beltran05a,boixo07a,luis07a,rey07a,boixo08a,choi07a,woolley08a}
is that the scaling of $\delta\gamma$ with $n$ depends on the probe
dynamics as expressed in $H$.  For an interaction that acts
independently on the probe constituents, the optimal measurement
precision scales as $1/n$, a scaling often called the ``Heisenberg
limit,'' as this was believed to be the best scaling allowed by the
Heisenberg uncertainty principle.  In contrast, a nonlinear
Hamiltonian that includes all possible $k$-body couplings gives an
optimal sensitivity that scales as $1/n^k$.  To achieve this requires
that the initial probe state be entangled.  If practical
considerations preclude initializing the probe in an entangled state,
sensitivity that scales as $1/n^{k-1/2}$ is possible using a probe
that is initially in a product
state~\cite{luis04a,beltran05a,luis07a,boixo08a,woolley08a}. Both of
these scalings can be achieved with separable measurements.

Practical interest in using nonlinear interactions for quantum
metrology comes from the fact that even with two-body couplings and
initial product states, it is possible to obtain a $1/n^{3/2}$
scaling for $\delta
\gamma$~\cite{luis04a,beltran05a,luis07a,boixo08a,choi07a,woolley08a}.
In some versions of this scheme, entanglement is generated during the
protocol that leads to the $1/n^{3/2}$ scaling.  We formulate here a
protocol that generates no entanglement among the probe constituents,
yet still achieves the $1/n^{3/2}$ scaling; in this protocol, it is
clearly the {\em dynamics\/} alone that leads to improvement over the
$1/n$ scaling.  Even though this Letter is mainly about improving on
the Heisenberg scaling, any experimental demonstration of a scaling
better than $1/n^{1/2}$ would be of considerable interest to the
metrology community.

A typical $k=2$ choice for a probe made of qubits is $H=J_z^{\,2}$,
where $J_z=\frac{1}{2}\sum_{j=1}^n Z_j$ is the $z$ component of the
``total angular momentum,''  with $Z_j$ being the Pauli $Z$ operator
for the $j\,$th qubit.  If we denote the eigenvectors of $Z$ by
$|0\rangle$ and $|1\rangle$, an optimal initial state is the
entangled ``half-cat" state, $(|0 \ldots 0 \rangle + |1 \ldots 1
\rangle) | 0 \ldots 0 \rangle) /\sqrt{2}$, where the first part of
the state refers to the first half of the qubits and the second part
to the rest. This state evolves to $(e^{-i\gamma tn^2/4}|0 \ldots
0\rangle+|1\ldots1\rangle)|0\ldots0\rangle)/\sqrt{2}$ after time $t$.
Measuring the product of Pauli $X$ operators on the first half of the
qubits gives a signal that oscillates in $\gamma$ with frequency
$tn^2/4$.  An estimate of $\gamma$ based on sampling from this signal
over $\nu$ trials leads to a measurement precision
$\delta\gamma=4/tn^2\sqrt\nu$~\cite{boixo07a}.

If the initial probe state is required to be a product state, an
optimal input state is of the form $e^{-iJ_y\beta}|0\rangle^{\otimes
n}=[\cos(\beta/2)|0\rangle+\sin(\beta/2)|1\rangle]^{\otimes n}$,
where $0<\beta\le\pi/2$. A measurement of $J_y$ after the probe has
evolved for a time $t$ under the $J_z^{\,2}$ Hamiltonian leads to a
measurement precision that scales as $1/tn^{3/2}\sqrt\nu$, provided
$\gamma t$ is small~\cite{boixo08a}.  This scaling applies for all
values of $\beta\ne\pi/2$, but the optimal sensitivity occurs for
$\beta=\pi/4$.  The restriction to small times arises because the
$J_z$ eigenstates in an expansion of the evolving state accumulate
phase shifts quadratic in $n$, leading to a ``phase dispersion'' that
after a short time renders it impossible to determine $\gamma$
optimally from a separable measurement such as that of $J_y$.

The $J_z^{\,2}$ Hamiltonian is entangling. The entanglement generated
during evolution from an initial product state and the phase
dispersion are two aspects of the same phenomenon.  One might think
that the generated entanglement and associated phase dispersion
somehow play a role in the enhanced $1/n^{3/2}$ scaling, but it would
normally be expected that the phase dispersion is best
avoided~\cite{boixo08a}.

The essential observation we make here is that if the $J_z^{\,2}$
Hamiltonian were replaced with one of the form $H =n J_z$, there
would be no phase dispersion and no generated entanglement.  An
$nJ_z$ Hamiltonian acts as a linear coupling whose strength is
proportional to $n$.  Physically, an $nJ_z$ coupling cannot arise
from a fundamentally linear coupling, as that would require the
coupling strength to be a function of the number of constituents in
the probe, but it can arise naturally from quadratic couplings to the
parameter.

With a pure $nJ_z$ interaction, the optimal initial product state is
$e^{-iJ_y\pi/2}|0\rangle^{\otimes
n}=[(|0\rangle+|1\rangle)/\sqrt{2}]^{\otimes n}$.  The state remains
unentangled at all times, evolving to $[(e^{-i\gamma
tn/2}|0\rangle+e^{i\gamma tn/2}|1\rangle)/\sqrt{2}]^{\otimes n}$. A
measurement of $J_x$ at time $t$ has expectation value $\langle
J_x\rangle =\frac{1}{2}n\cos\gamma tn$ and uncertainty $\Delta
J_x=\frac{1}{2}\sqrt n|\sin\gamma tn|$, leading to a measurement
precision $\delta\gamma=\Delta J_x/\sqrt\nu\,|d\langle
J_x\rangle/d\gamma|=1/tn^{3/2}\sqrt\nu$ after $\nu$ trials.  A
measurement of any other equatorial component of $\bm{J}$ achieves
the same sensitivity.  The enhanced scaling in a protocol that uses
an $nJ_z$ coupling and an initial product state is clearly due to the
dynamics alone, not to entanglement of the constituent qubits. These
results indicate that in quantum metrology, entanglement is important
only in providing an optimal initial state, which leads to an
improvement by a factor of $1/n^{1/2}$ over initial product states.

\begin{figure}
\resizebox{7.5 cm}{6.5cm}{\includegraphics{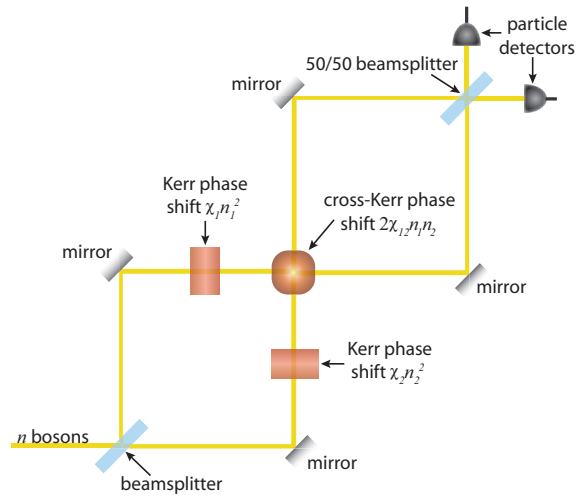}}
\caption{Nonlinear interferometer giving $J_z^2$ and $nJ_z$
couplings.  An incoming beam of $n$ bosons is split at a
beamsplitter, which puts each boson into an appropriate superposition
of being in the two arms (modes).  The two initial nonlinear phase
shifters produce Kerr phase shifts $\chi_1n_1^2$ and $\chi_2n_2^2$.
The phase shifter at the intersection of the beams produces a
cross-Kerr phase shift $2\chi_{12}n_1 n_2$.   The final 50/50
beamsplitter converts the required measurement of an equatorial
component of $\bm{J}$ into a measurement of $J_z$, i.e., a counting
of the difference of the numbers of particles in the two output
beams. The net effect of the nonlinear phase shifters is the same as
a probe Hamiltonian $\mathcal{H}$ acting for a time $t$,\vspace{.4em}
with\vspace{.4em}
\centerline{$\mathcal{H}t/\hbar=\chi_1n_1^2+\chi_2n_2^2+2\chi_{12}
n_1n_2$\hspace{10em}} \vspace{.4em}
\centerline{\hspace{2.4em}$=(\chi+\chi_{12})n^2/2+(\chi_1-\chi_2)nJ_z+2(\chi-\chi_{12})J_z^2\;,$}
where $\chi=\frac{1}{2}(\chi_1+\chi_2)$ is the average Kerr phase
shift. The first term in the second line produces an overall phase
shift and can be ignored.  The $nJ_z$ coupling comes from having
different Kerr phase shifters in the two arms; to eliminate the
$J_z^{\,2}$ interaction requires a cross-Kerr coupling
$\chi_{12}=\chi$.  Under these circumstances, we have
$\mathcal{H}=\hbar\gamma nJ_z$, with $\gamma t=\chi_1-\chi_2$. The
case $\chi_2=-\chi_1$ yields a pure $nJ_z$ coupling without a
compensating cross-Kerr phase shift.\label{fig0}}
\end{figure}

We are interested in investigating measurement protocols that use
both $J_z^{\,2}$ and $nJ_z$ interactions in systems of bosons that
can occupy two modes with creation operators $a_1^{\dagger}$ and
$a_2^{\dagger}$.  In the Schwinger representation, we have $J_z =
\frac{1}{2}(n_1-n_2)$ and $n=n_1+n_2$, where
$n_1=a_1^{\dagger}a_1^{\vphantom{\dagger}}$ and
$n_2=a_2^{\dagger}a_2^{\vphantom{\dagger}}$ are the numbers of
particles in the two modes.  The bosons we consider interact with one
another, but the interactions conserve particle number, so the system
has a nonzero chemical potential.  Our measurement protocols, for
both types of coupling, can be represented in terms of the
interferometer with nonlinear phase shifters depicted in
Fig.~\ref{fig0}.  In practical implementations, the interferometer
might be an optical or Ramsey interferometer or an interferometer
made up of coupled nanomechanical resonators~\cite{woolley08a}.

An $nJ_z$ coupling acts as a linear coupling with a coupling
strength proportional to $n$.  Thus the effect of decoherence on
our measurement protocol is the same as that on a linear protocol
with a product-state input.  In particular, decoherence that acts
independently on the probe particles does not change the $1/n^{3/2}$
scaling~\cite{boixo08a,woolley08a}.

We turn now to the problem of implementing the nonlinear
interferometer of Fig.~\ref{fig0} in a laboratory system of
considerable interest.  For this purpose~\cite{choi07a,rey07a} we
consider a two-mode Bose-Einstein
condensate (BEC) in which the $n$ atoms can occupy two internal
states (modes) labeled $|1\rangle$ and $|2\rangle$, which are
typically hyperfine levels.  The atoms that form the initial BEC are
all in the internal state $|1\rangle$.  In the mean-field
approximation, they all share the same spatial wave function
$\psi_n({\bm r})$, which is the $n$-dependent ground-state solution
of the Gross-Pitaevskii equation for a trapping potential $V({\bm
r})$ and a scattering term characteristic of internal state
$|1\rangle$.  An external field, playing the role of the first
beamsplitter in Fig.~\ref{fig0}, drives transitions between the two
internal states~\cite{hall98a}, resulting in every atom being in the
same superposition of the two internal states.  We assume that the
atomic collisions are elastic, so the only scattering channels are
$|1\rangle|1\rangle\rightarrow|1\rangle|1\rangle$,
$|2\rangle|2\rangle\rightarrow|2\rangle|2\rangle$, and
$|1\rangle|2\rangle\rightarrow|1\rangle|2\rangle$.  These have
amplitudes $g_{11}$, $g_{22}$, and $g_{12}$, where $g_{ij} =
4\pi\hbar^2a_{ij}/m$, with $a_{ij}$ being the $s$-wave scattering
length.  The effect we seek is the differential phase shift between
the two internal states due to their different scattering properties.
After some period of evolution, a second external field, playing the
role of the second beamsplitter in Fig.~\ref{fig0}, drives a $\pi/2$
pulse between the internal states.  A final measurement then
determines the population difference between the two internal states.
In the following we are interested in the BEC dynamics that occurs
between application of the external fields.

We assume that the two internal states are chosen so that both see
the same trapping potential $V$, which is a situation that can be
achieved in the laboratory.  Nonetheless, the spatial wave functions
corresponding to the two internal states will diverge because they
experience different scattering interactions.  The effect of the
scattering terms on the spatial wave functions becomes important at
the atom number $n_c$ where the scattering energy becomes comparable
to the total atomic kinetic energy.  For $n$ small compared to $n_c$,
the two spatial wave functions remain essentially the same, and
for $n$ much larger than $n_c$, the spatial changes, though they
become substantial, occur on a time scale longer than the phase
shifts of interest by a fractional power of $n/n_c$, which can be
around ten in laboratory experiments~\cite{mertes07a}.  We thus
neglect changes in the spatial wave functions, assuming that both
internal states retain the initial wave function $\psi_n({\bm r})$
for the duration of our proposed experiment.

With these assumptions the Hamiltonian for the two-mode
BEC~\cite{dalfovo99a,leggett01a} takes the form
\begin{equation}
\mathcal{H} =
\mathcal{H}_0 + \gamma_1\eta(n-1)J_z + \gamma_2\eta J_z^{\,2}\;,
\end{equation}
where $\eta = \int d\bm{r}\,|\psi_n(\bm{r})|^4$, $\gamma_1 =
\frac{1}{2}(g_{11}-g_{22})$, $\gamma_2=g-g_{12}$, and
$g=\frac{1}{2}(g_{11}+g_{22})$ (notice that $\gamma_1$ and $\gamma_2$
do not have units of frequency).  The only effect of the Hamiltonian
$\mathcal{H}_0 = nE_0 + \frac{1}{4}(g+g_{12})\eta n^2 -
\frac{1}{2}g\eta n$, where $E_0$ is the single-particle kinetic plus
trap potential energy corresponding to $\psi_n$, is to introduce an
overall phase, and thus $\mathcal{H}_0$ can be ignored.  We assume
$n$ is large enough that we can replace $n-1$ with $n$ in
$\mathcal{H}$.

In a harmonically trapped BEC, the repulsive scattering interactions
cause the single-particle ground-state wave function $\psi_n$ to
spread as the number of particles increases.  This effect appears in
the BEC Hamiltonian in the factor $\eta$, which is inversely
proportional to the effective volume occupied by the ground-state
wave function. The $n$ dependence of $\eta$ gives the coupling
strength a dependence on $n$ that must be included in our analysis of
the precision in estimating $\gamma_{1,2}$.

When the number of atoms is small compared to $n_c$, the total
kinetic energy far exceeds the scattering energy, resulting in a
ground-state wave function that is independent of $n$.  In a
three-dimensional harmonic trap with ground-state half-width $s$, the
total kinetic energy is $\sim n(\hbar^2/ms^2)$, and the scattering
energy is $\sim n^2(g_{11}/s^3)$ (for atoms in internal
state~$|1\rangle$), giving $n_c\sim s/a_{11}$.  Typical values of
$a_{11}\sim 10\,\mbox{nm}$ and $s\sim10\,\mbox{$\mu$m}$ give
$n_c\sim1\,000$. Hence, for a condensate composed of tens to a few
hundred or so atoms, $\eta$ does not depend significantly on $n$,
implying a scaling of $1/n^{3/2}$ in such small BECs.

In large harmonically trapped BECs, with $n\gg n_c$, $\eta$ acquires
an $n$ dependence that defeats the desire to improve on $1/n$
scaling.  Strategies for dealing with this include using traps with
harder walls than a harmonic trap and working with BECs confined to
fewer than three dimensions.  To assess these strategies, we compute
the $n$ dependence of $\eta$ when the BEC is trapped in $d$
longitudinal dimensions by a spherically symmetric potential
$V=\frac{1}{2}kr^q$ and is tightly confined in the remaining $D=3-d$
transverse dimensions by a harmonic potential.  The longitudinal trap
is characterized by the hardness parameter $q$ and the half-width of
its (bare) ground-state wave function, $R_0=(\hbar^2/mk)^{1/(q+2)}$,
for which a typical value might be $R_0\sim 10\,\mbox{$\mu$m}$.  The
tight transverse potential is characterized by its resonant frequency
$\omega_0$ and the half-width $s=(\hbar/2m\omega_0)^{1/2}$ of its
ground-state wave function, for which a typical value for a tight
trap would be $s\sim100\,\mbox{nm}$.

There are now two critical atom numbers. The first,
$n_L=(R_0/a_{11})(s/R_0)^D$, occurs when the scattering energy is
comparable to the longitudinal kinetic energy. As $n$ increases from
$n_L$, the ground-state wave function spreads in the longitudinal
dimensions, its size growing as $R\sim R_0(n/n_L)^{1/(q+d)}$. The
second critical atom number, $n_T=(s/a_{11})(R_0/s)^{d(q+2)/q}$,
arises when the scattering energy becomes as large as the transverse
kinetic energy (and thus does not apply when $d=3$), at which point
the longitudinal extent of the wave function is $R_T$.  The
corresponding atomic number density, $n_T/s^D
R_T^{\,d}\sim1/a_{11}s^2\sim10^{16}\,\mbox{cm}^{-3}$, is somewhat
above the upper limit on number density set by three-body scattering
losses.  Thus we need only consider atom numbers smaller than $n_T$.

For atom numbers between $n_L$ and $n_T$, a reasonable approximation
to the ground-state solution of the Gross-Pitaevskii equation is
obtained by using the Gaussian ground state of width $s$ in the
transverse dimensions and using the Thomas-Fermi approximation for
the longitudinal wave function~\cite{das02a}.  In this approximation
we find
\begin{equation}
\label{eta}
\eta=\frac{2q}{2q+d}\frac{\lambda}{ng_{11}}
=\frac{\alpha_{q,d}}{s^D R_0^{\,d}}\left(\frac{n_L}{n}\right)^{d/(d+q)}\;,
\end{equation}
where $\lambda=\mu-\frac{1}{2}D\hbar\omega_0$ is the longitudinal
part of the chemical potential $\mu$ and $\alpha_{q,d}$ is a
geometric factor of order unity that depends on $d$ and $q$, but not
on $n$.  The $n$ dependence of $\eta$ implies an effective coupling
strength that scales as $n^{\xi-1/2}$, where $\xi=(d+3q)/2(d+q)$.
The precision of estimating $\gamma_1$ or $\gamma_2$ thus scales as
$1/n^\xi$.

For a three-dimensional BEC trapped in a harmonic potential, the
measurement precision scales as $1/n^{9/10}$, worse than the
Heisenberg scaling, but still better than $1/n^{1/2}$.  To achieve
super-Heisenberg scalings requires a trapping potential that is
harder than a harmonic potential or else working with a one- or
two-dimensional BEC.  For $d=2$, a BEC trapped in a harmonic
potential matches the $1/n$ scaling, and a one-dimensional harmonic
BEC betters it, achieving a $1/n^{7/6}$ scaling.  A $d$-dimensional
BEC achieves super-Heisenberg scaling when the hardness parameter $q$
exceeds $d$.  The limit of large $q$ corresponds to a trap with hard
walls and extent $2R_0$ and has $\xi=3/2$ regardless of $d$.  For a
one-dimensional BEC, an alternative to hard caps is to use a ring
geometry.

A good candidate for implementing the generalized metrology protocol
is a BEC made of\ \rbd\ atoms.  Atoms in the hyperfine level
$|F=1;M_F=-1\rangle=|1\rangle$ are trapped and cooled to form a BEC,
and then a Raman or microwave-driven transition is used to create a
superposition of $|1\rangle$ and the hyperfine level~\cite{hall98a,matthews98a}
$|F=2;M_F=1\rangle=|2\rangle$.  The
$s$-wave scattering lengths $a_{11}$, $a_{22}$, and $a_{12}$ are
nearly degenerate for \rbd, with ratios $\{ a_{22} : a_{12} : a_{11}
\} = \{0.97 : 1 : 1.03 \}$. These values imply that $\gamma_2 =
\frac{1}{2}(g_{11}+g_{22})-g_{12}$ is essentially zero for this
scheme, meaning that a \rbd~BEC can realize the generalized quantum
metrology protocol with a pure $n^{\xi-1/2}J_z$ coupling.  The
optimal initial state for this protocol has all atoms in an equally
weighted superposition of $|1\rangle$ and $|2\rangle$.  The quantity
that is estimated is proportional to $\gamma_1 =
\frac{1}{2}(g_{11}-g_{22})$, which, though small, is nonzero for the
scattering lengths in \rbd.

Loss of atoms from the trap is an important decoherence mechanism,
mainly due in our protocol to inelastic spin-exchange collisions
(exchange of atoms with the thermal cloud that is present around any
realistic BEC is negligible and can be ignored).  A chief advantage
of using protocols that do not rely on entanglement is that loss of
atoms does not affect the sensitivity scaling, although it does
generally degrade the sensitivity.  In the case of spin-exchange
collisions, the decoherence can be modeled in terms of a parameter
$\Gamma\eta/2$, which we can estimate using data
from~\cite{mertes07a} and the assumption that $|1\rangle$ and $|2
\rangle$ have the same spatial wave function.  This estimate gives
$\Gamma/2\gamma_1\sim 1/26$, implying that we can perform a
measurement of $\gamma_1$ before inelastic collisions have a
significant impact.

A final issue is that the number of atoms in a BEC is not known to
arbitrary precision, as we have assumed up till now.  We propose to
determine $n$ by counting the number of atoms in both internal states
at the output of our protocol.  A determination of $n$ with a
fractional error of $\Delta n/n\sim 0.01$, which is within current
capabilities, would be sufficient for the purpose of demonstrating an
enhanced scaling with $n$, provided the measurement time is kept
short enough that the nonlinear phase shift is much smaller than
$n/\Delta n$.  We note that if $\Delta n$ is bigger than $\sqrt n$,
the chief practical advantage of the $nJ_z$ interaction is obviated,
since the requirement on measurement time is as strict as or stricter
than that set by phase dispersion in a $J_z^2$ protocol. Even so, the
ability of the $nJ_z$ coupling to achieve enhanced scalings with no
generated entanglement remains an important theoretical objective.
Moreover, decoherence is likely to limit the measurement time more
severely than either phase dispersion or number uncertainty.

We have shown that it is possible to achieve measurement precision
that scales better than $1/n$ by using the dynamics generated by
nonlinear Hamiltonians.  The pure $nJ_z$ scheme introduced here does
not use quantum entanglement at any stage to achieve the enhanced
scaling. Early experiments to test our scheme in BECs are likely to
focus on demonstrating enhanced scaling in the estimation of some
combination of atomic scattering lengths. To be useful, however, our
scheme must be adapted to measuring external fields that modulate the
atomic scattering properties.  One possibility is to use a \csm~BEC
with optical trapping of the $|F=3;M_F=3\rangle $
state~\cite{weber03}, which has a very broad Feshbach resonance at
$8\,\mbox{G}$, which makes the scattering lengths very sensitive to
the strength of an external magnetic field~\cite{leo00,chin00}.  This
suggests that our scheme might be used for ultra-high precision
magnetometry.

The authors thank I.~H.~Deutsch and G.~J. Milburn for advice on the
theory and practice of atomic BECs.  This work was supported in part
by the US Office of Naval Research (Grant No.~N00014-07-1-0304), the
Australian Research Council's Discovery Projects funding scheme
(Project No.~DP0343094), and the National Nuclear Security
Administration of the US Department of Energy (Contract
No.~DE-AC52-06NA25396).  STF was supported by the Perimeter Institute
for Theoretical Physics; research at Perimeter is supported by the
Government of Canada through Industry Canada and by the Province of
Ontario through the Ministry of Research~\&\ Innovation.

\bibliography{becbib}

%%
%% TABLES
%%
%% If there are any tables, put them here.
%%

\end{document}